\documentclass[aps,prl,twocolumn,showpacs,superscriptaddress,floatfix,nofootinbib]{revtex4}
\usepackage{graphicx}
\usepackage{amsmath}
\usepackage{epsfig}
\usepackage{helvet}
\usepackage{amssymb}

\newcommand{\be}{\begin{equation}}
\newcommand{\ee}{\end{equation}}
\newcommand{\bea}{\begin{eqnarray}}
\newcommand{\eea}{\end{eqnarray}}

\newcommand{\tr}{\mbox{tr}}
\newcommand{\bra}[1]{\mbox{$\langle #1 |$}}
\newcommand{\ket}[1]{\mbox{$| #1 \rangle$}}
\newcommand{\braket}[2]{\mbox{$\langle #1  | #2 \rangle$}}
\newcommand{\proj}[1]{\mbox{$|#1\rangle \!\langle #1 |$}}

\def\tr{ \mbox{tr}}

\begin{document}

\title{
Classical simulation of quantum many-body systems with a tree tensor
network}
\author{Y.-Y.~Shi}
\affiliation{Department of Electrical Engineering and Computer
Science, University of Michigan, Ann Arbor, MI 48109, USA}
\author{L.-M.~Duan}
\affiliation{Department of Physics, University of Michigan, Ann
Arbor, MI 48109, USA}
\author{G.~Vidal} \affiliation{School of Physical Sciences, the University of
Queensland, QLD 4072, Australia}

\date{\today}

\begin{abstract}
We show how to efficiently simulate a quantum many-body system with
tree structure when its entanglement is bounded for any bipartite
split along an edge of the tree. This is achieved by expanding the
{\em time-evolving block decimation} simulation algorithm for time
evolution from a one dimensional lattice to a tree graph, while
replacing a {\em matrix product state} with a {\em tree tensor
network}. As an application, we show that any one-way quantum
computation on a tree graph can be efficiently simulated with a
classical computer.
\end{abstract}


\maketitle

\noindent {\bf Introduction.--} The underlying physical laws
describing classical and quantum systems are quite different, which
makes it generally hard to efficiently simulate a quantum many-body
evolution with a classical computer. However, if the quantum
evolution has certain restrictions, an efficient classical
simulation may be possible
\cite{1,2,4,Vid,VC_ent,PEPS,Shi,Tobi}. Indeed, efficient simulation
algorithms are known for several restricted quantum circuit models
\cite{1,2}, while a general connection between entanglement and the
classical simulatability of quantum lattice models has been unveiled
\cite {4,Vid,VC_ent}. In one spatial dimension (1D), for instance,
the state of a quantum chain with a limited amount of entanglement
between any bipartition along the chain can be efficiently described
using a {\em matrix product state} (MPS) \cite{MPS}.  This is
exploited by the {\em density matrix renormalization group} (DMRG)
algorithm \cite{7} to find the ground state of the chain and by the
{\em time-evolving block decimation} (TEBD) method \cite{Vid} to
simulate an evolution in time. Furthermore, {\em projected
entangled-pair states} (PEPS) have been introduced as an extension
of MPS to simulate 2D systems \cite{PEPS}.

In this work we consider the simulation of a quantum many-body
system with a restricted amount of entanglement according to a tree
structure. As in the 1D and 2D approaches mentioned above, we
express the $d^n$ complex amplitudes $c_{i_1\cdots i_n}$ of the
state $\ket{\Psi}$ of $n$ {\em qudits} (or $d$-level quantum
systems),
\begin{equation}\label{eq:psi}
    \ket{\Psi} = \sum_{i_1=1}^d \cdots\sum_{i_n=1}^d c_{i_1\cdots i_n} \ket{i_1}\otimes \cdots \otimes
    \ket{i_n},
\end{equation}
in terms of a network of tensors \cite{tensor_network}, but
specialize to the case where this tensor network (TN) has tree
structure. Given a tree TN, we explain (i) how to simulate the
response of the system to local operations and classical
communication (LOCC), that is to generic manipulation of individual
qudits, including adaptive unitary transformations and measurements
and (ii) how to simulate time evolution. The latter is achieved by
extending the TEBD algorithm \cite{Vid}, originally proposed to
simulate 1D quantum lattices, so that it applies to a much broader
class of states and physical situations. As in the 1D case, the key
to an efficient simulation is that the amount of entanglement in the
system remains sufficiently bounded during the time evolution.

From the perspective of theoretical computer science, our results
imply that one-way quantum computation with a tree-graph cluster
state can be efficiently simulated with a classical computer.
One-way quantum computation with cluster states \cite{BR}, an
interesting alternative to the quantum circuit model, had previously
been shown to be universal for quantum computation in a 2D lattice
\cite{RB} but classically simulatable in a 1D lattice \cite{11}. By
extending the classical simulatability result to tree cluster
states, we further sharpen the boundary between the complexities for
classical and quantum computation.

From the perspective of computational physics, our results provide
an algorithm both to find the ground state and to simulate time
evolution in complex quantum systems with tree structure (see also
\cite{previous}), such as dendrimers --a class of highly branched
polymers \cite{dendrimers}. This algorithm, based solely on tensor
multiplications and singular value decompositions, also offers new
ways to simulate 1D systems with long-range interactions.

\vspace{1mm}

\begin{figure}
  \includegraphics[width=8cm]{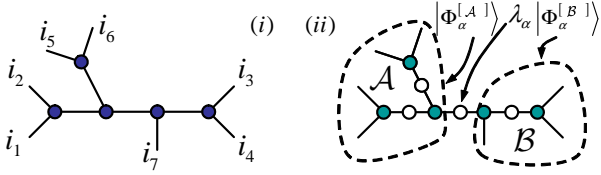}\\
\caption{({\em i}) A tree TN for state $\ket{\Psi}$ of $n=7$ qudits.
Each open index labels an orthonormal single-qudit basis, cf Eq.
(\ref{eq:psi}). Each vertex has three edges and corresponds to a
tensor with three indices. Pairs of tensors are connected in the
network through a {\em shared} index, over which there is an
implicit summation. ({\em ii}) Canonical form of the previous tree
TN. For each bipartition, sub-trees ${\cal A}$ and ${\cal B}$
describe Schmidt bases, see Eq. (\ref{eq:Schmidt}). An empty circle
on top of an edge represents a set of Schmidt coefficients weighting
the corresponding index. }\label{fig:canonical}
\end{figure}

\noindent {\bf Canonical form of a tree TN.--} We use a TN with $n$
{\em open} indices and tree structure, see Fig.~(\ref{fig:canonical}),
to express the $d^n$ complex coefficients of
the $n$-qudit state $\ket{\Psi}$. More specifically, we consider a
tree network made of tensors with only 3 indices each
\cite{three_enough}, the network therefore containing $n-2$ tensors.
An index connecting two tensors divides the network into two
sub-trees ${\cal A}$ and ${\cal B}$ and the $n$ qudits into two
disjoint sets. We use the term {\em bipartition} to refer only to
such divisions. The rank of an index is the number of values it
takes. In what follows, $\chi$ denotes the largest rank among all
indices in the network. Notice that the tree TN depends on
$O(n\chi^3)$ complex coefficients.

The Schmidt decomposition of $\ket{\Psi}$ according to bipartition
${\cal A}\!:{\cal B}$ reads
\begin{equation}\label{eq:Schmidt}
\ket{\Psi} = \sum_{\alpha=1}^{\chi_{({\cal A}\!:{\cal B})}}
\lambda_{\alpha} \ket{\Phi^{[{\cal A}]}_{\alpha}} \otimes
\ket{\Phi^{[{\cal B}]}_{\alpha}},
\end{equation}
where $\braket{\Phi_{\alpha}^{[{\cal A}]}}{\Phi_{\alpha'}^{[{\cal
A}]}}= \braket{\Phi_{\alpha}^{[{\cal B}]}}{\Phi_{\alpha'}^{[{\cal
B}]}}= \delta_{\alpha\alpha'}$, $\sum_\alpha (\lambda_\alpha)^2 = 1$
and where the Schmidt rank satisfies $\chi_{({\cal A}\!:{\cal B})}
\leq \chi$.  We next introduce the {\em canonical form} of a tree
TN, which also consists of tensors with three indices but where each
index $\alpha$ shared by two tensors carries weights, see
Fig.~(\ref{fig:canonical}).

\vspace{1mm}

\noindent {\bf Definition.} A tree TN is in the {\em canonical form}
for bipartition ${\cal A}\!:{\cal B}$ if ($i$) the weights on the
connecting index $\alpha$ correspond to the Schmidt coefficients
$\{\lambda_{\alpha}\}$ and ($ii$) the subtrees ${\cal A}$ and ${\cal
B}$ describe a set of Schmidt bases $\{\ket{\Phi^{[{\cal
A}]}_{\alpha}}\}$ and $\{\ket{\Phi^{[{\cal B}]}_{\alpha}}\}$. A tree
TN is in the canonical form if it is so for all bipartitions.

\vspace{1mm}

\noindent {\bf Theorem 1.} {\em The canonical form of an $n$-qudit
tree TN $n$ can be obtained with ${\cal O}(n\chi^4)$ basic
operations.}

\begin{figure}
  \includegraphics[width=8cm]{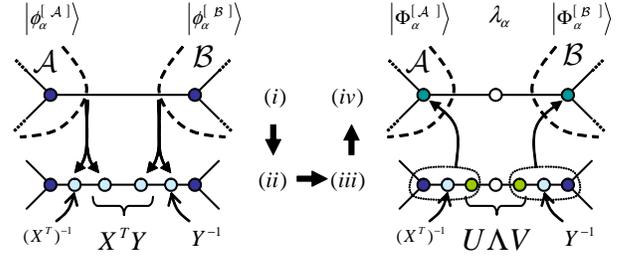}\\
\caption{({\em i}) Initial decomposition, Eq.
(\ref{eq:non-Schmidt}). ({\em ii}) Insertion of $(X^{T})^{-1}X^{T}$
and $YY^{-1}$, projectors onto the subspaces generated by
$\{\ket{\hat{e}_{\tau}}\}$ and $\{\ket{\hat{f}_{\eta}}\}$. ({\em
iii}) $X^{T}Y$ is replaced with its singular value decomposition
$U\Lambda V$, compute in $O(\chi^3)$ time. ({\em iv}) The Schmidt
decomposition, Eq. (\ref{eq:Schmidt}), is obtained after contracting
some indices ($O(\chi^4)$ time).}\label{fig:ortho}
\end{figure}

\noindent {\em Proof:} Given a bipartition ${\cal A}:{\cal B}$,
$\ket{\Psi}$ can be written as
\begin{equation}\label{eq:non-Schmidt}
\ket{\Psi} = \sum_{\alpha} \ket{\phi_{\alpha}^{[{\cal A}]}}\otimes
\ket{\phi_{\alpha}^{[{\cal B}]}},
\end{equation}
where the sets of states $\{\ket{\phi_{\mu}^{[{\cal A}]}}\}$ and
 $\{\ket{\phi_{\mu}^{[{\cal B}]}}\}$ for subtrees ${\cal A}$ and
${\cal B}$ may not be orthonormal. Our goal is to turn Eq.
(\ref{eq:non-Schmidt}) into the Schmidt decomposition
(\ref{eq:Schmidt}). Let
\begin{equation}\label{eq:scalar}
    M = \tilde{X}D\tilde{X}^{\dagger},  ~~~~~~ M_{\alpha\alpha'} \equiv
\braket{\phi^{[{\cal A}]}_{\alpha'}}{\phi^{[{\cal
    A}]}_{\alpha}},
\end{equation}
be the spectral decomposition of the matrix $M$ of scalar products
in ${\cal A}$, where $\tilde{X}^{\dagger}\tilde{X} =
\tilde{X}\tilde{X}^\dagger = I,$ and $D_{\tau\tau'} =
\delta_{\tau\tau'} d_{\tau}$. Then the vectors
\begin{equation}\label{eq:ortho}
    \ket{\hat{e}_{\tau}} \equiv \frac{1}{\sqrt{d_{\tau}}} \sum_{\alpha}
    (\tilde{X}^{\dagger})_{\tau\alpha} \ket{\phi_{\alpha}^{[{\cal A}]}},
    ~~~~~~~~ d_{\tau} > 0,
\end{equation}
form an orthonormal set. We define $X \equiv \tilde{X}\sqrt{D}$.
Analogous considerations in ${\cal B}$ lead to an orthonormal set
$\{ \ket{\hat{f}_{\eta}}\}$ and a matrix $Y$. Eq.
(\ref{eq:non-Schmidt}) can be rewritten as
\begin{equation}\label{eq:non-Schmidt2}
\ket{\Psi} = \sum_{\tau \eta} (X^{T}Y)_{\tau \eta}
\ket{\hat{e}_{\tau}}\otimes \ket{\hat{f}_{\eta}}.
\end{equation}
From the singular value decomposition (SVD) of $X^T Y$,
\begin{equation}\label{eq:svd}
    (X^T Y) = U \Lambda V,
\end{equation}
where $U^{\dagger}U=VV^{\dagger}=I$ and $\Lambda$ is a diagonal
matrix, we obtain the Schmidt coefficients $\lambda_{\alpha} =
\Lambda_{\alpha\alpha}$ and bases
\begin{equation}\label{eq:Schmidt_all}
 \ket{\Phi_{\alpha}^{[{\cal A}]}} = \sum_{\tau} (U^T)_{\alpha\tau}
\ket{\hat{e}_{\tau}},~~~
 \ket{\Phi_{\alpha}^{[{\cal B}]}} =
\sum_{\eta} V_{\alpha\eta} \ket{\hat{f}_{\eta}}.
\end{equation}
All the above steps, summarized in Fig.~(\ref{fig:ortho}), take
$O(\chi^4)$ time and are repeated for each of the $n-3$ bipartitions
of the tree TN. It is not hard to see that the scalar product
matrices $M$ for all edges can be computed sequentially in an
appropriate order also in time $O(n\chi^4)$ time.



\vspace{1mm}

\begin{figure}[h]
  \includegraphics[width=7cm]{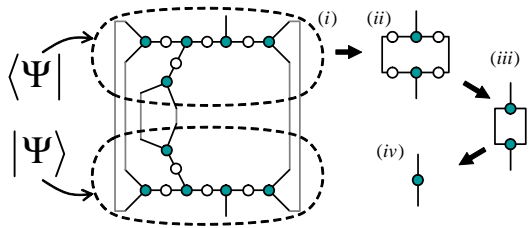}\\
\caption{A single-qudit reduced density matrix is computed by
contracting the TN in ({\em i}), that represents $\proj{\Psi}$ with
a partial trace over all qudits but one. This TN reduces to the TN
in ({\em ii}) thanks to the orthogonality of the Schmidt bases. In
({\em iii}) the weights have been absorbed into the three-index
tensors ($O(d\chi^2)$ operations) and in ({\em iv}) the remaining
shared indices have been contracted ($O(d^2\chi^2)$ operations).
}\label{fig:rho1}
\end{figure}

\noindent {\bf Manipulating a tree TN.--} We now describe some basic
tasks involving a tree TN, that is assumed to be in the canonical
form. First, we can compute the reduced density matrix for just one
or two qudits. Fig.~(\ref{fig:rho1}) shows how to proceed for one
qudit.

\vspace{1mm}

\begin{figure}[h]
  \includegraphics[width=8cm]{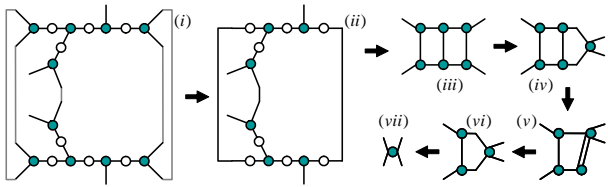}\\
\caption{The reduced density matrix for two qudits separated by
$m=3$ tensors is computed by contracting the TN in ({\em i}). This
reduces to the TN in ({\em ii}) thanks to the orthogonality of the
Schmidt bases. In ({\em iii}) the Schmidt weights have been absorbed
into the tensors ($O(d\chi^2)$ operations), which in turn are
re-organized on a line made of $m=3$ columns. Networks ({\em iv}) to
({\em vi}) illustrate how to reduce by one the number of columns
($O(d^2\chi^4)$ operations). By iteration, a single tensor ({\em
vii}) with four open indices is obtained.}\label{fig:rho2}
\end{figure}

\noindent {\bf Theorem 2.} {\em A two-qudit reduced density matrix
can be computed with $O(md^2\chi^4)$ basic operations, where $m$ is
the number of tensors in the path that connects the two qudits in
the tree TN.} [see Fig.~(\ref{fig:rho2})]

Second, we can arbitrarily relocate a qudit within the tree TN. This
is achieved by swapping the index corresponding to this qudit with
other indices in the tree.

\vspace{1mm}

\begin{figure}[h]
  \includegraphics[width=7cm,height=13mm]{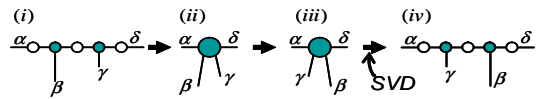}
\caption{In order to exchange the position of indices $\beta$ and
$\gamma$ in ({\em i}), we contract the corresponding tensors
(including all neighboring weights) into the four-legged tensor in
({\em ii}). After swapping the two indices, we split the resulting
tensor ({\em iii}) through a SVD (of a matrix of the dimension
$d\chi\times d\chi$ or $d\chi\times \chi^2$) that costs $O(d^2\chi^4)$ basic
operations. The singular values define the new weights of the
central index, whose initial rank $\chi$ may have increased to at
most $d\chi$. The old weights for lateral indices are detached from
the two tensors to leave the resulting tree TN in the canonical
form.} \label{fig:swap}
\end{figure}

\noindent {\bf Theorem 3.} {\em Swapping a qudit index of a tensor
with an index of a neighboring tensor can be achieved with
$O(d^2\chi^4)$ basic operations.} [See Fig.~(\ref{fig:swap})].

Third, we can update the tree TN after a unitary gate $U$ has acted
either on one qudit or on two neighboring qudits. Unitarity
preserves the orthogonality of the Schmidt bases for most
bipartitions, and as a result the update process involves changing
only one or two tensors. When $U$ acts on one qudit or on two qudits
that are connected to the same three-legged tensor, that tensor
simply absorbs the gate through index contraction:
\begin{figure}[h]
  \includegraphics[width=5cm,height=11mm]{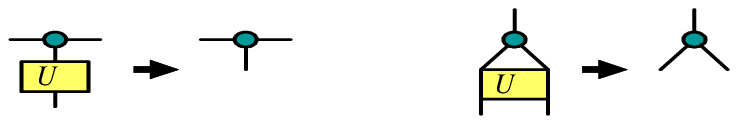}\\
\label{fig:Usimple}
\end{figure}

\vspace{1mm}

\begin{figure}[h]
  \includegraphics[width=5cm,height=13mm]{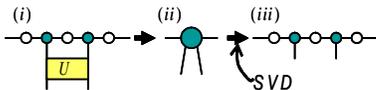}\\
\caption{ The tensor subnetwork in ({\em i}) is contracted into the
four-legged tensor in ({\em ii}), which is then decomposed into
({\em iii}) by using a SVD (of a $d\chi\times d\chi$ matrix)
that takes $O(d^3\chi^3)$ basic
operations. Notice that, as in Fig.~(\ref{fig:swap}), special
attention is paid to first absorbing and then detaching the weights
of the lateral indices. This guarantees that the resulting tree TN
is in its canonical form. }\label{fig:svd}
\end{figure}

\noindent {\bf Theorem 4.} {\em Consider a two-qudit gate U acting
on a pair of open indices of two tensors that are nearest neighbors
in the network. The tree TN can be updated by replacing these two
tensors, at a cost of $O(d^3\chi^3)$ basic operations.} [See
Fig.~(\ref{fig:svd})].

\vspace{1mm}

\noindent {\bf Efficient simulation with a tree TN.--} All the above
manipulations require computational time (and space as well)
that scales at most
linearly in the number of qudits $n$ and as a small power of the
maximal Schmidt rank $\chi$. Therefore, in those systems where the amount of
entanglement across all relevant bipartitions, as characterized by
$\chi$, scales at most polynomially in $n$, such manipulations can
be implemented efficiently. This opens up a number of simulation
possibilities.

For instance, we can simulate the response of the system to
arbitrary local manipulation. Recall that LOCC manipulation can be
decomposed as an adaptive sequence of generalized local measurements
mapping pure states into pure states. Let ${\cal E}$ denote one such
measurement on a qudit, as characterized by a set of operators
$\{E_{r}\}$, where $r$ labels the measurement outcome. Outcome $r$
occurs with probability $p_r= \bra{\Psi}E_{r}^{\dagger} E_r
\ket{\Psi}$, in which case the state of the system becomes
$\ket{\Psi_r} = E_r\ket{\Psi}/\sqrt{p_r}$. To simulate ${\cal E}$,
first we randomly draw an outcome $r$ according to the probability
$p_r = \tr [E_r\rho^{(1)}E_{r}^{\dagger}]$, computed from the
reduced density matrix $\rho^{(1)}$ of the qudit to be measured.
Then a tree TN for $\ket{\Psi_r}$ is obtained from that of
$\ket{\Psi}$ by simply absorbing operator $E_r$ into it. The new
maximal Schmidt rank $\chi_r$ satisfies $\chi_{r} \leq \chi$.

An implication of the above result is that one-way quantum
computation on a tree can be efficiently simulated. This follows
from the fact that a tree-graph cluster state has a very simple tree
TN representation, with $\chi = 2$, while the manipulations involved
in a one-way computation consists of LOCC.

The simulation of a time evolution according to a two-body
Hamiltonian $H = \sum_{i,j} h_{ij}$ is also possible. As in the case
of a 1D system \cite{Vid}, we expand the evolution operator
$V=\exp(-iHt)$ into a series of two-qudit unitary gates $U$ using a
Suzuki-Trotter expansion. But now, for each of these gates, we first
bring the indices of the two qudits together using Theorem 3, then
absorb $U$ into the tree TN using Theorem 4, and finally bring the
qudit indices back into their initial position. This generalizes the
TEBD algorithm \cite{Vid} from 1D systems to a generic tree TN. With
minimal modifications to deal with non-unitary gates, The TEBD
algorithm can also be used to simulate an evolution in imaginary
time according to $V'=\exp(-Ht)$. In this way we can compute the
ground state of $H$, provided $H$ has a finite gap $\Delta>0$ in its
spectrum. Recall that the expectation value of local observables,
such as the energy $E = \bra{\Psi}H\ket{\Psi}$ or two-point
correlators, can be computed from two-qubit reduced density
matrices, which we can obtain using Theorem 2.

As discussed in \cite{Vid} for 1D systems, the use of a canonical
form related to the Schmidt decomposition has an important advantage
in situations where $\chi$ is too large. There $\ket{\Psi}$
may still be reasonably {\em approximated} by a tree TN with a
smaller effective $\tilde{\chi}$. For a given bipartition ${\cal
A}:{\cal B}$, the best approximation to $\ket{\Psi}$ is obtained by
{\em truncating} decomposition (\ref{eq:Schmidt}) so as to retain
only the $\tilde{\chi}$ terms with greatest weight
$\lambda_{\alpha}$. Thus, the canonical form of the tree TN is ideal
to implement optimal bipartite truncations.

\begin{figure}[h]
  \includegraphics[width=8cm]{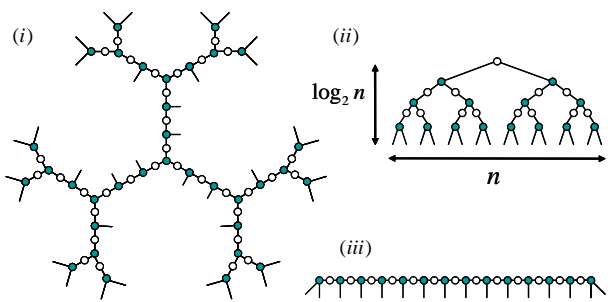}\\
\caption{({\em i}) Tree TN for a system with tree structure, such as
a dendrimer. ({\em ii}) Two qudits in a binary tree TN are connected
through at most $O(\log(n))$ tensors. ({\em iii}) Two qudits in a
linear TN are typically connected through $O(n)$
tensors.}\label{fig:fractal}
\end{figure}

A tree TN may be used to simulate a many-body system with a genuine
tree structure (determined e.g. by the interaction pattern) as it is
the case of dendrimers \cite{dendrimers}. But it can also be used to
simulate a 1D system with long-range interactions, including
periodic boundary conditions. Notice that when using a MPS (that is,
a linear TN) to represent a 1D system, the number of tensors in the
path connecting two random qudits scales as $O(n)$ and,
consequently, so does the computational time to simulate a gate
acting on those qudits. Instead, in a binary tree TN, see
Fig.~(\ref{fig:fractal}), the path connecting any two qudits contains at
most $O(\log n)$ tensors. This shortens the time required to
simulate that gate by a significant factor $\log(n)/n$.

We conclude by noticing that in this paper we have explored the
most general extension of the TEBD algorithm. Indeed, it appears
that a tree TN
--- not having closed loops --- is the most general TN to which we can
associate a Schmidt decomposition to each of its indices, a
fundamental ingredient of the simulation algorithm.

Y.-Y. thanks A. Kitaev for introducing to him the general concept of
tensor networks. L.-M. thanks S. Barrett, P. Kok, and F. Verstraete
for helpful discussions. This work was supported by NSF under Awards
0323555, 0347078, and 0431476, the ARDA under ARO contracts, and the
A.~P.~Sloan Fellowship.


\begin{thebibliography}{99}
\bibitem{previous} Ref. \cite{Caley} computes the exact ground state on a Caley tree for very specific spin
models, while attempts to extend DMRG to tree structures have been
reported in \cite{treeDMRG}. Ref. \cite{dendrimers} considers DMRG
on a tree, but where the Hilbert space dimension only grows linearly
in the system size $n$.
\bibitem{three_enough} A tensor with more than three indices can be
decomposed into a tree TN made of tensors with three indices each
(for instance, through a sequence of SVDs).
\end{thebibliography}
\end{document}